\documentclass{aa}
\usepackage{graphicx}
\usepackage{txfonts}

\usepackage{natbib,twoopt}
\usepackage[breaklinks=true]{hyperref} 

\hypersetup{
  colorlinks=true,   
  urlcolor=blue,     
  linkcolor=blue     
}
\bibpunct{(}{)}{;}{a}{}{,} 

\makeatletter
\newcommand{\bibnote}[2]{\global\@namedef{#1note}{#2}}
\newcommand{\biblink}[2]{\global\@namedef{#1link}{#2}}
\makeatother

\makeatletter
  \protected\def\stonyslink{%
     \def\hyper@linkstart##1##2{}\let\hyper@linkend\@empty}
  \newcommandtwoopt{\citeads}[3][][]{%
   \href{http://adsabs.harvard.edu/abs/#3}%
        {\stonyslink \citealp[#1][#2]{#3}}
   \biblink{#3}{\href{http://adsabs.harvard.edu/abs/#3}{ADS}}}
 \newcommandtwoopt{\citepads}[3][][]{%
   \href{http://adsabs.harvard.edu/abs/#3}%
        {\stonyslink \citep[#1][#2]{#3}}
   \biblink{#3}{\href{http://adsabs.harvard.edu/abs/#3}{ADS}}}
 \newcommandtwoopt{\citetads}[3][][]{%
   \href{http://adsabs.harvard.edu/abs/#3}%
        {\stonyslink \citet[#1][#2]{#3}}
  \biblink{#3}{\href{http://adsabs.harvard.edu/abs/#3}{ADS}}}
 \newcommandtwoopt{\citeyearads}[3][][]{%
   \href{http://adsabs.harvard.edu/abs/#3}%
        {\stonyslink \citeyear[#1][#2]{#3}}
   \biblink{#3}{\href{http://adsabs.harvard.edu/abs/#3}{ADS}}}
\makeatother

\newcommand*\psj{PSJ}

\begin{document} 

   \title{Trojan asteroids and co-orbital dust ring of Venus}
\author{Yang-Bo Xu\inst{1,2}
	\and
	Lei Zhou\inst{1,2}
	\and
	Christoph Lhotka\inst{3}
	\and 
	Li-Yong Zhou\inst{1}\fnmsep\inst{2}
	\and
	Wing-Huen Ip\inst{4}
}
\authorrunning{Xu et al.}
\offprints{Zhou L.-Y., \email zhouly@nju.edu.cn}
\institute{School of Astronomy and Space Science, Nanjing University, 163 Xianlin Avenue, Nanjing 210023, China
	\and
	Key Laboratory of Modern Astronomy and Astrophysics in Ministry of Education, Nanjing University, China
	\and
	Space Research Institute, Austrian Academy of Sciences, Schmiedlstrasse 6, 8042 Graz, Austria
	\and
	Institute of Astronomy, National Central University, Jhongli, Taoyuan City 32001, Taiwan
    }
   \date{}

  \abstract
   {The long-standing co-orbital asteroids have been thought to be the possible source of zodiacal dust ring around the orbit of Venus, but inconsistent conclusions on the orbital stability thus existence of Venus Trojans are found in literature.}
  {We present in this paper a systematic survey of the orbital stability of Venus Trojans, taking into account the dynamical influences from the general relativity and the Yarkovsky effect. }
  {The orbits of thousands of fictitious Venus Trojans are numerically simulated. Using the method of frequency analysis, their orbital stabilities and the dynamical mechanisms behind are described in detail. The influences of general relativity and of Yarkovsky effect, which were either neglected or oversimplified previously, are investigated by long-term numerical simulations. }
  {The stability maps on the $(a_0,i_0)$ plane and $(a_0,e_0)$ plane are depicted, and the most stable Venus Trojans are found to occupy the low-inclination horseshoe orbits with low eccentricities. The resonances that carve the fine structures in the stability map are determined.  The general relativity decreases the stability of orbits by little but the Yarkovsky effect may drive nearly all Venus Trojans out of the Trojan region in a relatively short time. }
  {The Venus Trojans have poor orbital stability and cannot survive the age of the Solar system. The zodiacal dust ring found around the orbit of Venus is more likely a sporadic phenomenon, as the result of temporarily capture into the 1:1 mean motion resonance of dust particles produced probably from passing comets or asteroids, but not Venus Trojans. }
  
   \maketitle
%

\section{Introduction}\label{sec:intro}

In the circular restricted 3-body problem (R3BP), the equilateral triangular Lagrange equilibrium points $L_4$ and $L_5$ are dynamically stable for all planets in the solar system \citepads[see e.g.][]{1996CeMDA..65..149E,1999ssd..book.....M}. For a long time these solutions had been considered only as of theoretical interest until \citetads{1907AN....174...47W} discovered the asteroid (588) Achilles librating around the $L_4$ point of Jupiter. Such asteroids were named by their discoverers after mythic characters in the Greek narratives about the Trojan War. From then on, it has become a custom to call a celestial object librating around $L_4$ or $L_5$ of any planet a Trojan. By now several thousand Trojans have been found orbiting around the orbit of Jupiter. And small numbers of Trojan asteroids have also been found around the Earth, Mars, Uranus and Neptune. 

Sharing the same orbit with a planet means a Trojan asteroid has the same orbital period as the host planet, i.e. they are in fact in the 1:1 mean motion resonance (MMR). The Trojan dynamics is always of special interest and has been studied theoretically and numerically for most of the planets in the solar system. A comprehensive theoretical analysis in R3BP can be found e.g. in \citetads{1967torp.book.....S}. A formal long-periodic solution for Trojans in R3BP was constructed by \citetads{1977AJ.....82..368G,1978CeMec..18..259G}.
Based on the 3-dimensional elliptic R3BP, an approximate analytical theory was developed by \citetads{1981CeMec..24..377E,1984CeMec..34..435E,1988CeMec..43..303E}, and the perturbations from other planets were taken into account by considering Jupiter's orbit as a secularly changing ellipse \citepads{1996IAUS..172..171E}.

\citetads{1992AJ....104.1641M} explored the evolution of Trojan-type asteroidal orbits with numerical simulations, and with the advancement of computation ability, the numerical methods have been applied widely. Nearly a century after the discovery of Jupiter Trojans, the detection of the first Mars Trojan in 1990 \citepads{1990BAAS...22.1357B} evoked many researches on the dynamical stability \citepads{1994AJ....107.1879M, 1999ApJ...517L..63T, 2005Icar..175..397S} and the origin of Mars Trojans \citepads{2013Icar..224..144C, 2015Icar..252..339C, 2017NatAs...1E.179P}.
A decade later, the first Neptune Trojan was discovered \citepads{2003MPEC....A...55P}. The origin and long-term stability of Neptune Trojans have been thoroughly studied \citepads{2007MNRAS.382.1324D,2009MNRAS.398.1217Z,2011MNRAS.410.1849Z,2009MNRAS.398.1715L}.
Another decade later, the discovery and confirmation of the first Earth Trojan \citepads{2011ApJ...731...53M, 2011Natur.475..481C} aroused the studies on the long-term stability of Earth Trojans \citepads{2012A&A...541A.127D, 2013CeMDA.117...91M, 2019A&A...622A..97Z}. The first Uranus Trojan was discovered in 2011 \citepads{2013MPEC....F...19A}, before which their dynamical stabilities had been deeply studied \citepads{2002Icar..160..271N,2003A&A...410..725M} and \citetads{2010CeMDA.107...51D} postulated their existence at low and also high inclinations, and this issue is thoroughly clarified recently in \citetads{2020A&A...633A.153Z}.

As for Venus Trojans (VTs for short), the observation is challenging because of their close proximity to the Sun. A few searches in the Trojan region of Venus report no detection of  stable VTs \citepads{2020PSJ.....1...47P, 2020AJ....159...70Y} so far. However, there are a few studies on the dynamical stability and possibility of their existence. 
\citetads{2000MNRAS.319...63T} found stable VT orbits of both tadpole and horseshoe types over the integration timescale of 100\,Myr and restricted the stable region with respect to inclinations to below $16^\circ$. 
\citetads{2005AJ....130.2912S} presented a stability analysis of tadpole orbits over gigayear timescale. They applied Laskar's frequency map analysis to derive dynamical maps, from the most stable region of which they integrated 30 chosen orbits to investigate the long-term stability. Under the pure gravitational model of the solar system (consisting of the Sun and eight planets), all the 30 test particles escaped after 1.2\,Gyr. Including the Yarkovsky effect of a certain strength, the lifespan decreases to 400\,Myr. Thus the authors concluded that any population of primordial VTs of tadpole orbits would have disappeared by now. Using direct numerical integrations under the same gravitational model, \citetads{2012MNRAS.426.3051C}, however, find that a substantial number of VTs of both tadpole and horseshoe orbits appear to be stable over 1\,Gyr. 

Recently, this controversy is re-aroused by \citetads{2019ApJ...873L..16P}. Their main concern is the origin of Venus's zodiacal dust ring rather than the long-term stability of VTs, but they find that a group of hypothetical VTs may serve as the most essential source of the dust ring. They enlarge the number of test particles to 10\,000 including both tadpole and horseshoe orbits, and extend the integration to the age of the Solar system (4.5\,Gyr). Using the same pure gravitational model, they find that 8.2\% of the test particles remain stable for the whole simulation, which they think confirms the notion that primordial VTs could still exist today. However, the Yarkovsky effect, has been ignored in this research, though its importance had been briefly shown by \citetads{2005AJ....130.2912S}. 

The first evidence for the presence of the dust ring associated with Venus is attributed to the measurements of the Venera 9 and 10 spacecraft \citepads{1979P&SS...27..951K} and by measurements of the Helios mission \citepads{2007A&A...472..335L}, but the conclusive measurements were finally obtained with STEREO \citepads[]{2013Sci...342..960J, 2017Icar..288..172J} and by the Parker Solar Probe \citepads{2021ApJ...910..157S}.  The retention time of dust particles in the 1:1 MMR is less than 0.1--1\,Myr before they are driven out by the drag induced by the so-called Poynting-Roberston effect \citepads{2014Icar..232..249K}. Therefore, the dust ring, if it was not a temporary phenomenon, must be continuously filled up with new particles either from the erosion of primordial VTs as \citetads{2019ApJ...873L..16P} suggest, or from the zodiacal dust cloud. It should be carefully examined. 

In one word, the dynamical structure in the Trojan region of Venus is not clearly known, and the mechanisms that produce and sustain the dust ring associated with Venus is still ambiguous. We also note that the general relativity has not been included in previous studies, although it could introduce considerable long-term influence to the motion of objects in the vicinity of the Sun. Both the stability of VTs and its implication on the existence of the dust ring around Venus deserve an investigation in  greater details.

Aiming at clarifying whether the primordial VTs serve as the major source of dust particles around the orbit of Venus, we present here an investigation on the dynamics of VTs with frequency analysis method and numerical simulations. The rest of this paper is organized as follows. We introduce our model and methods in Sect.~\ref{sec:modmtd}. The orbital stability of VTs is studied in Sect.~\ref{sec:stab}. Then the effect of general relativity will be examined in Sect.~\ref{sec:gryark}, followed by the investigation of the influence of Yarkovsky effect. Finally, we conclude this paper in Sect.~\ref{sec:disc}.

\section{Model and Method}\label{sec:modmtd}

\subsection{Dynamical model} \label{subsec:model}

The dynamical model we adopt consists of the Sun, all planets in the solar system and massless fictitious VTs. We treat the Earth and Moon as a whole in their barycentre. The initial planetary configuration at epoch of JD 245\,7400.5 is taken from JPL HORIZONS system\footnote{\url{ssd.jpl.nasa.gov/horizons.cgi}} \citepads{1996DPS....28.2504G}. To explore the dynamical behaviour of VTs with different inclinations, we initialize the test particles with a grid on the $(a_0,i_0)$ plane. The initial semi-major axis ranges from 0.71533 AU to 0.73133 AU with a step of $10^{-4}$ AU while the inclination is sampled from $0^\circ$ to $60^\circ$ at interval of $1^\circ$. Since there is no remarkable dynamical asymmetry between two triangular Lagrange points, we just focus on the $L_4$ point and initially place VTs there with $\omega=\omega_2+60^\circ$, where $\omega$ is the argument of perihelion\footnote{The subscripts `1' to `8' denote the planets Mercury to Neptune respectively, as usual.}. For other angular elements including the longitude of the ascending node $\Omega$ and mean anomaly $M$, we set the same values as Venus. Hence the initial resonant angle
\begin{equation}
\sigma=\lambda-\lambda_2\,,
\end{equation}
where $\lambda=\omega+\Omega+M=\varpi+M$ is the mean longitude, is always $60^\circ$ for all VTs. The initial eccentricity $e$ is set to be the same value as Venus $e=e_2=0.00675$, 
as we did in our previous work \citepads{2009MNRAS.398.1217Z,2012A&A...541A.127D,2019A&A...622A..97Z,2020A&A...633A.153Z}. 
This low eccentricity value is representative partly because the stable regions for Trojan asteroids around different planets are found to locate mainly at low eccentricities \citepads[e.g.][]{2005Icar..175..397S,2005AJ....130.2912S,2011MNRAS.410.1849Z}. The dependence on eccentricity will be further explored later in Section \ref{sec:stab}.

\subsection{Analysis methods}
The whole system is integrated up to $\sim$10\,Myr via a Lie-series integrator \citepads{1984A&A...132..203H} for stability analysis. We use the similar analysis method as we did in \citetads{2009MNRAS.398.1217Z,2011MNRAS.410.1849Z,2019A&A...622A..97Z,2020A&A...633A.153Z}. 
An on-line low-pass digital filter is applied to the output of the integration to remove the short-period terms \citepads{1995A&A...303..945M,2002Icar..158..343M}, and then the fast Fourier transform (FFT) is applied to the filtered data to derive the frequency spectra.

The spectral number (SN) is defined as the number of peaks above a specific noise level in a power spectrum. The peaks are found under the criterion that their amplitudes are larger than the amplitudes of nearest four frequency points in the power spectrum. The critical amplitude is set to 1\% of the highest peak, and the amplitudes below which are considered as the noise.
The SN indicates the regularity of an orbit and generally reflects the long-term stability. The superiority and reliability of SN as the stability indicator for Trojans have been confirmed in previous work \citetads{2009MNRAS.398.1217Z, 2011MNRAS.410.1849Z, 2019A&A...622A..97Z, 2020A&A...633A.153Z}.

To understand the structure seen in the dynamical map, we determine the locations of resonances that dominate the orbital behaviour of VTs. For this purpose, the proper frequencies are precisely calculated via the method developed by \citetads{1990Icar...88..266L,1993PhyD...67..257L}, which basically is an iterative scheme  to search for the maximum of the amplitude of 
\begin{equation}
	\Psi(\omega)=\frac{1}{2T}\int_{-T}^{T}\!f(t)\mathrm{e}^{-\mathrm{i}\omega{t}}\chi(t)\,d t\,,
\end{equation}
where $f(t)$ is a quasi-periodic function obtained numerically over time span $[-T,T]$, $\omega$ is the target frequency and the weight function $\chi(t)$ is the Hanning window $\chi(t)=1+\cos(\pi{t}/T)$. And a semi-analytical method is then used (see Sect.~\ref{subsec:fma}) to determine the location resonances. 

The integrator package \emph{Mercury6} \citepads{1999MNRAS.304..793C}, with some modifications to include the general relativity and Yarkovsky effect, is also adopted in our long-term numerical simulations of VTs' motion to the age of the Solar system. 

\section{Dynamical Stability}\label{sec:stab}

\subsection{Dynamical map}\label{subsec:dymap}
Making use of the SN of $\cos\sigma$, we construct the dynamical maps. A step-size of $h=0.01$\,yr is adopted to integrate the orbits, and after the low-pass filter the outputs are given in every $2^{13}\cdot h=81.92$\,yr. Totally $2^{17}$ lines of outputs are recorded for the frequency analysis, so that the total integration time is $T\approx 1.07\times10^7$\,yr. This time is longer than all the precession periods of planets (except the nodal precession of Jupiter, see Table~\ref{tab:ffss}.), thus these results can be used to reveal the involved secular mechanisms (for details of choosing integration step-size and output intervals, please refer to \citetads{2009MNRAS.398.1217Z}). The Nyquist frequency is thus $6.10\times10^{-3}\,2\pi\,$yr$^{-1}$ while the resolution is $9.35\times10^{-8}\,2\pi\,$yr$^{-1}$. In addition, results in a shorter time $T/8\approx 1.34\times10^6$\,yr will be also used to obtain some details of mechanisms of short time scale. 

\begin{figure}[htbp]
	\centering
	\includegraphics[width=9cm]{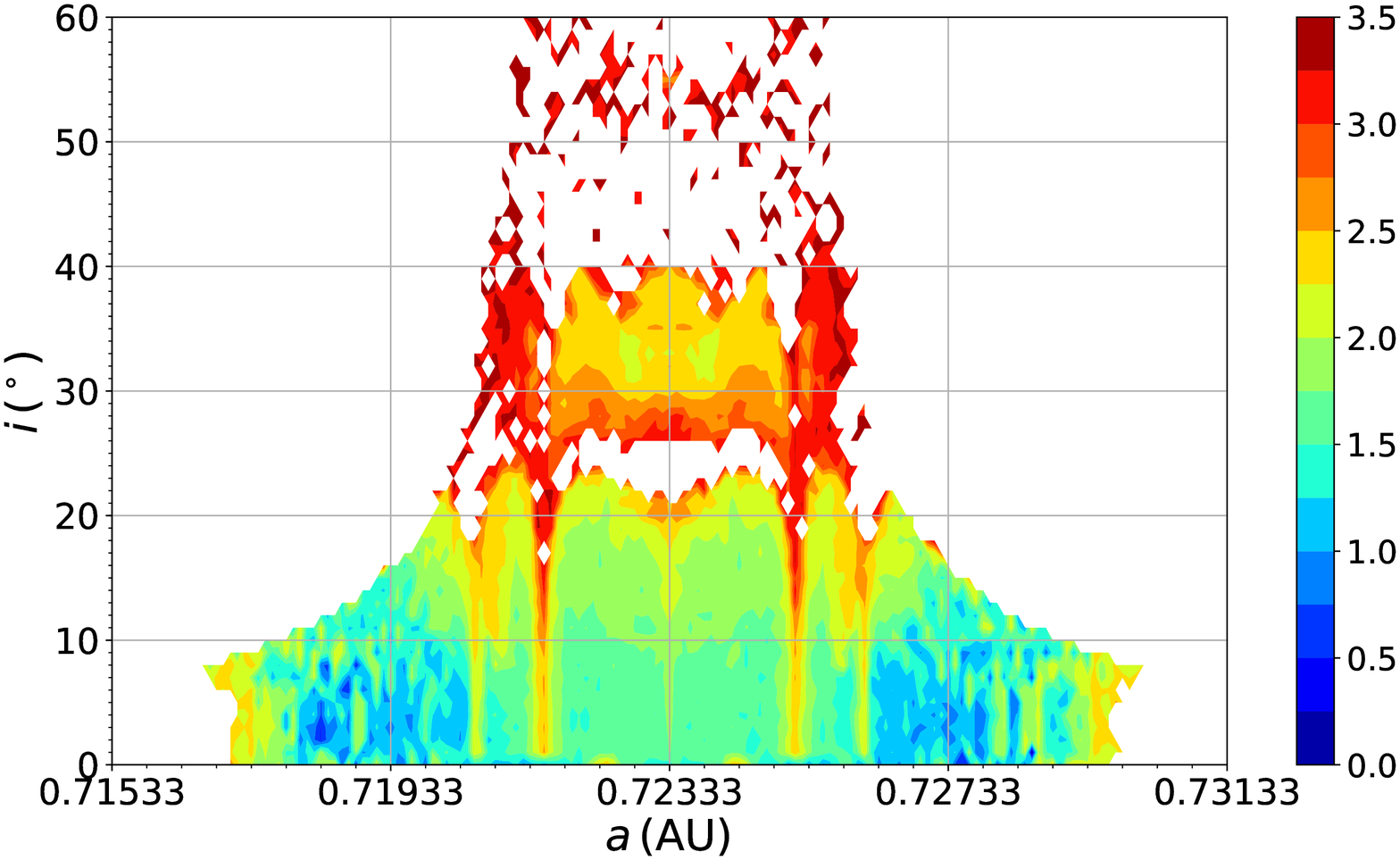}
	\includegraphics[width=9cm]{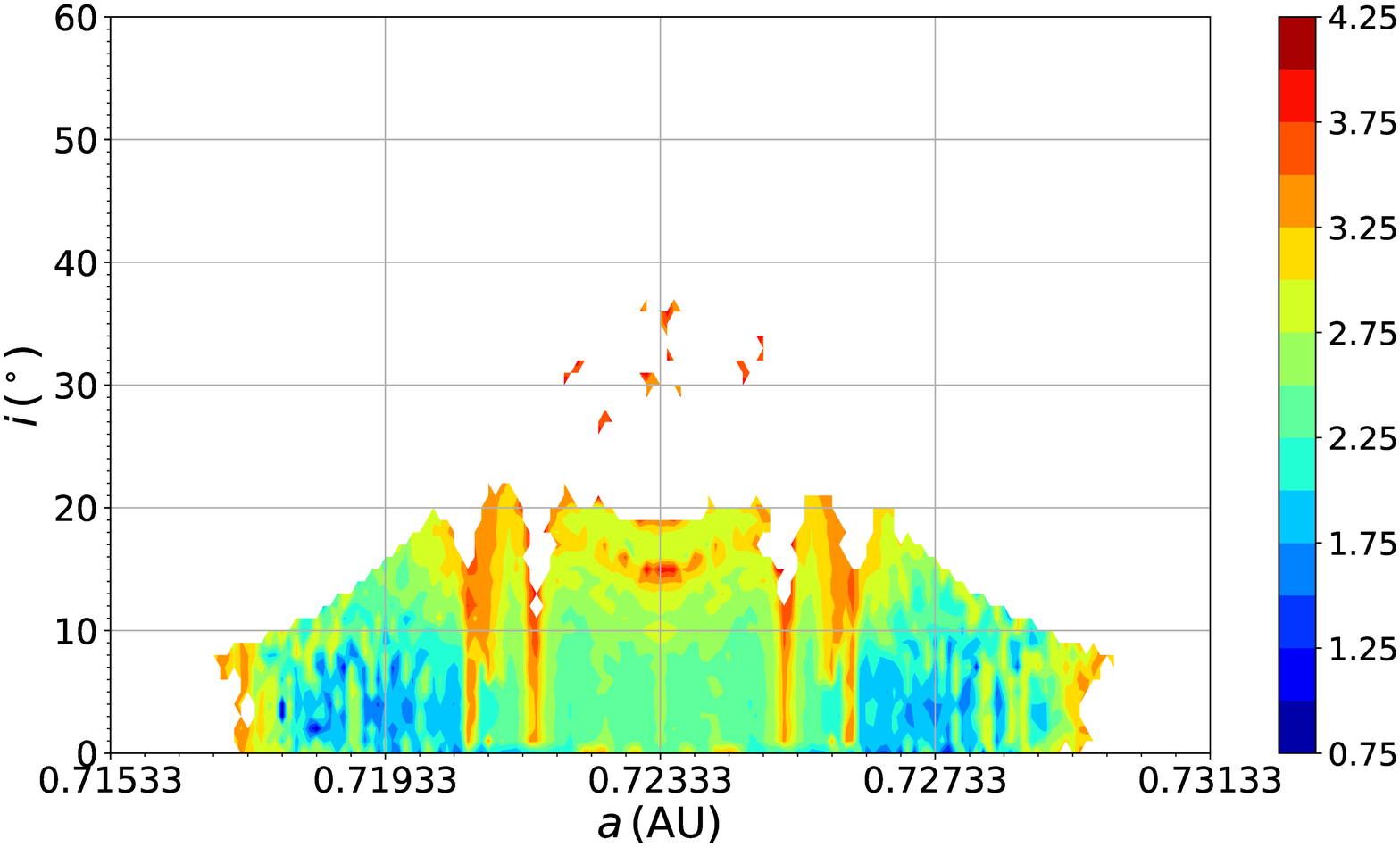}
	\caption{Dynamical maps around the $L_4$ point of Venus on the $(a_0,i_0)$ plane for the integration time of $1.34\times10^6$\,yr (\textit{top}) and $1.07\times10^7$\,yr (\textit{bottom}). The colours indicate the base-10 logarithm of SN of $\cos\sigma$ for orbits surviving the integration time.}
	\label{fig:dynmap}
\end{figure}

The dynamical maps on the $(a_0,i_0)$ plane for these two integration times are shown in Fig.~\ref{fig:dynmap}. The blue orbits have the smallest SN, which means the greatest stability and the red ones are the opposite. On the maps we exclude the orbits that escape from the co-orbital region of Venus\footnote{We define the escape as dissatisfying $|a-a_2| \le 0.02\,{\rm AU}$ after several test simulations.} within the integration time. 

Overall, the dynamical maps are similar to the one for Earth Trojans \citepads[see Fig.~1 of][]{2019A&A...622A..97Z}. Symmetrically distributed around the libration centre at $a=0.72333$\,AU, the most regular orbits occupy the low-inclination region of $i\lesssim 10^\circ$, the stable island extends up to $i\approx 20^\circ$ and is interrupted by an unstable gap around $i\approx 24^\circ$. A much less regular region can be seen in between $28^\circ\sim 40^\circ$. Compared to the Earth Trojans, the island at moderate inclinations ($\sim$30$^\circ$) is much less stable, and it disappears in the longer integration (bottom panel of Fig.~\ref{fig:dynmap}). Furthermore, the low-inclination stable island below $i=20^\circ$ is solid, not like the dynamical maps for Earth Trojans in which instability cavities exist at low inclinations. 

A pair of instability strips stand vertically at $0.72333\pm0.0018$ AU. They are the separatrix of tadpole and horseshoe orbits (see Fig.~\ref{fig:resweb}), and they could bring in chaos. The edge of the stability region is defined by the Hill radius of Venus ($\sim0.00676$\,AU) and on both sides of the region there lie the most stable VTs on horseshoe orbits. Conclusively, there do exist some stability regions where the stable VTs, especially those on horseshoe orbits, could reside, even for the age of the solar system (see Sect.~\ref{subsec:longt}).

For a short integration time ($1.34\times10^6$\,yr), more details can be seen on the dynamical map for moderate- and high-inclined orbits. The instability gap around $24^\circ$ is proven to be involved with the apsidal secular resonances with the Earth ($\nu_3$\footnote{By convention, we denote the linear apsidal secular resonance with the $j^{\rm th}$ planet by $\nu_j$ and the linear nodal secular resonance by $\nu_{1j}$.}) and Mars ($\nu_4$) \citepads{2002MNRAS.334..241B,2005AJ....130.2912S}. 

As time goes on, some mechanisms of longer time scale start to work, such as those related to the red belt around $15^\circ$ on the dynamical map for $1.07\times10^7$\,yr. We suppose that the orbits over there should escape in the near future.

\begin{figure}[htbp]
	\centering
	\resizebox{\hsize}{!}{\includegraphics{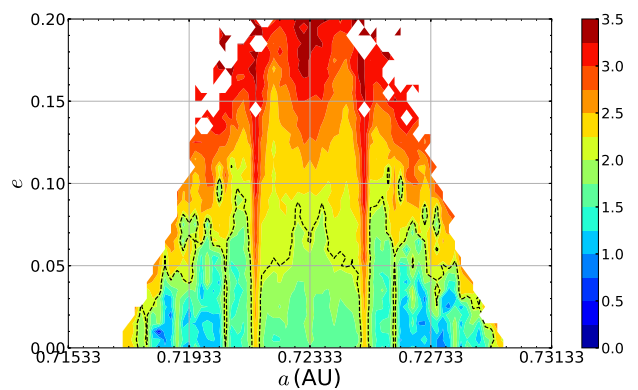}}
	\caption{Dynamical maps around the $L_4$ point of Venus on the $(a_0, e_0)$ plane for  integration time of $1.34\times10^6$\,yr. The colours indicate the base-10 logarithm of SN of $\cos\sigma$ for orbits surviving the integration time. The dashed contour curve indicates $\log_{10}\text{SN}=2.06$ (see text for details).}
	\label{fig:dynmap_e}
\end{figure}

In calculating Fig.~\ref{fig:dynmap}, the initial eccentricity is fixed at $e_0=0.00675$, so it provides a representative collection of orbits at low eccentricity region where the majority of stable orbits reside. To check the dependence of stability on initial eccentricity, we calculate the SN on the $(a_0, e_0)$ plane, this time with fixed initial inclination at $5^\circ$. The dynamical map obtained from orbit integrations of $1.34\times10^6$\,yr is shown in Fig.~\ref{fig:dynmap_e}. 

The stable regions are located at low eccentricities area, with the most stable orbits (in blue) occupying the region $e_0\lesssim 0.05$. The dashed contour in Fig.~\ref{fig:dynmap_e} encloses those orbits with $\text{SN}<10^{2.06}$, with which the orbit most probably will survive 4.5Gyr in the pure gravity model (see discussion in Sect.~\ref{subsec:longt}). The vertical structures seen in Fig.~\ref{fig:dynmap}, e.g. the unstable strips corresponding to the separatrix between tadpole and horseshoe orbit, and the strips around $a_0=0.72053, 0.72613$\,AU, can also be seen in Fig.~\ref{fig:dynmap_e}, implying that they arise due to mechanisms associated with semimajor axis (mean motion). 

In the low eccentricity region in Fig.~\ref{fig:dynmap_e}, we note that the feature of the dynamical map varies mainly along the semimajor axis (horizontally) but not along the eccentricity (vertically), i.e. the dynamics does not sensitively depend on eccentricity in this low eccentricity region. In fact, this phenomenon can be observed too in \citetads[][Figs.\,1 and 2 therein]{2005AJ....130.2912S}, where we also find in the fitting polynomials for proper frequencies ($g, s$) that the eccentricity terms contribute much less than the inclination and the semimajor axis variation. More evident dependence on eccentricity appear only in either high inclination ($\gtrsim 20^\circ$) or high eccentricity ($\gtrsim 0.17$) regions \citepads[see Fig.\,4 in][]{2005AJ....130.2912S}. But these regions are for very unstable orbits, thus not our concern in this paper. Therefore, below in the rest of this paper, we focus on the dynamical maps on $(a_0,i_0)$ plane and keep the setting of initial eccentricity as $e_0=0.00675$.

\subsection{Frequency analysis}\label{subsec:fma}

Basically, it is the secular resonances that carve the dynamical maps of Trojan asteroids of terrestrial planets \citepads[see e.g.][]{2002MNRAS.334..241B,2019A&A...622A..97Z}. To find out the secular resonances responsible for the structures in the phase space, we conduct a frequency analysis method.

The proper frequencies of VTs are distinguishable from the forced frequencies because they are dependent on the orbital elements. Hence we can pick them out from the frequency spectra. For details of determining the proper frequencies, please refer to \citetads{2009MNRAS.398.1217Z}. After obtaining the proper frequencies, we fit their dependencies on the orbital elements of $a_0$ and $i_0$. Surely, the proper frequencies should also depend on eccentricity $e_0$. However, with fixed $e_0$, we focus on the low eccentricity region, and will not analyse the eccentricity dependence in this paper.

We calculate and fit the proper rates of the perihelion precession $g=g(a_0,i_0)$ and of the ascending node precession $s=s(a_0,i_0)$, which correspond to the proper frequencies of $e\cos\varpi$ and $i\cos\Omega$  of VTs, respectively. Since the orbital behavior of tadpole and horseshoe orbits are different from each other, their proper frequencies follow different dependencies on the orbital elements. The piecewise quintic functions are adopted to deal with frequencies in these two regimes.

The secular resonances can be depicted by searching the integral domain for the combination of parameters ($p,q,p_j,q_j$) of the equation
\begin{equation} \label{eq:res}
pg+qs+\sum_{j=1}^{8}(p_jg_j+q_js_j)=0\,,	
\end{equation}
where $g_j$ and $s_j$ represent the precession rates of the perihelion and ascending node of the $j^{\rm th}$ planet. The proper frequencies of both the planets (see Table~\ref{tab:ffss}) and VTs are computed from the output of the $1.07\times10^7$\,yr integration. The d’Alembert rule requires $p+q+\sum_{j=1}^{8}(p_j+q_j)=0$ and $q+\sum_{j=1}^8q_j$ must be even. The value of $|p|+|q|+\sum_{j=1}^{8}\left(|p_j|+|q_j|\right)$ is defined as the degree of the secular resonance.

\begin{table}[htbp]
	\centering
	\caption{Proper frequencies of the planets in the solar system. The data is computed from the $1.07\times10^7$ yr integration using the method developed by \citetads{1990Icar...88..266L} except for $s_5$, which we take from \citetads{1989A&A...210..313N}.}
	\begin{tabular}{crr|crr}
		\hline\hline
		& Period~~~ & Freq.~ &  & Period~~~ & Freq.~ \\
		\hline
		$g_1$ & 248,447.20 & 40.25 & $s_1$ & $-235,349.49$ & $-42.49$ \\
		$g_2$ & 176803.39 & 56.56 & $s_2$ & $-189,789.33$ & $-52.69$ \\
		$g_3$ & 74,900.76 & 133.51 & $s_3$ & $-68,728.52$ & $-145.50$ \\
		$g_4$ & 72,542.62 & 137.85 & $s_4$ & $-73,115.45$ & $-136.77$ \\
		$g_5$ & 305,157.16 & 32.77 & $s_5$ & $-129,550,000.$ & $-0.08$ \\
		$g_6$ & 47,056.61 & 212.51 & $s_6$ & $-49,127.98$ & $-203.55$ \\
		$g_7$ & 419,463.09 & 23.84 & $s_7$ & $-431,778.93$ & $-23.16$ \\
		$g_8$ & 1,930,501.9 & 5.18 & $s_8$ & $-1,876,172.61$ & $-5.33$ \\
		\hline
	\end{tabular}
	\tablefoot{The periods are given in years and the frequencies are given in $10^{-7}\,2\pi\,{\rm yr}^{-1}$.}
	\label{tab:ffss}
\end{table}

\begin{figure*}[htbp]
	\centering
	\includegraphics[width=9cm]{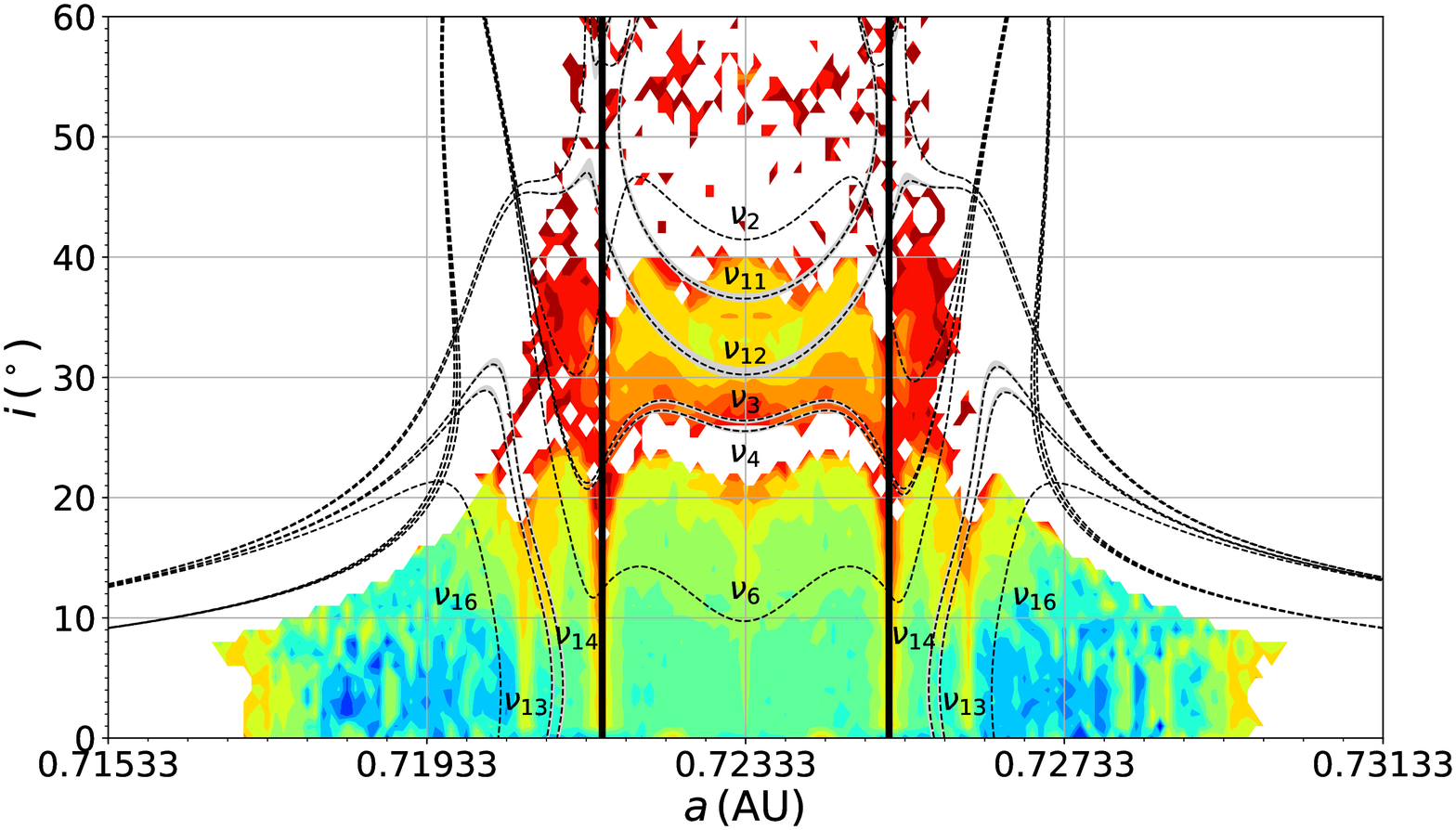}
	\includegraphics[width=9cm]{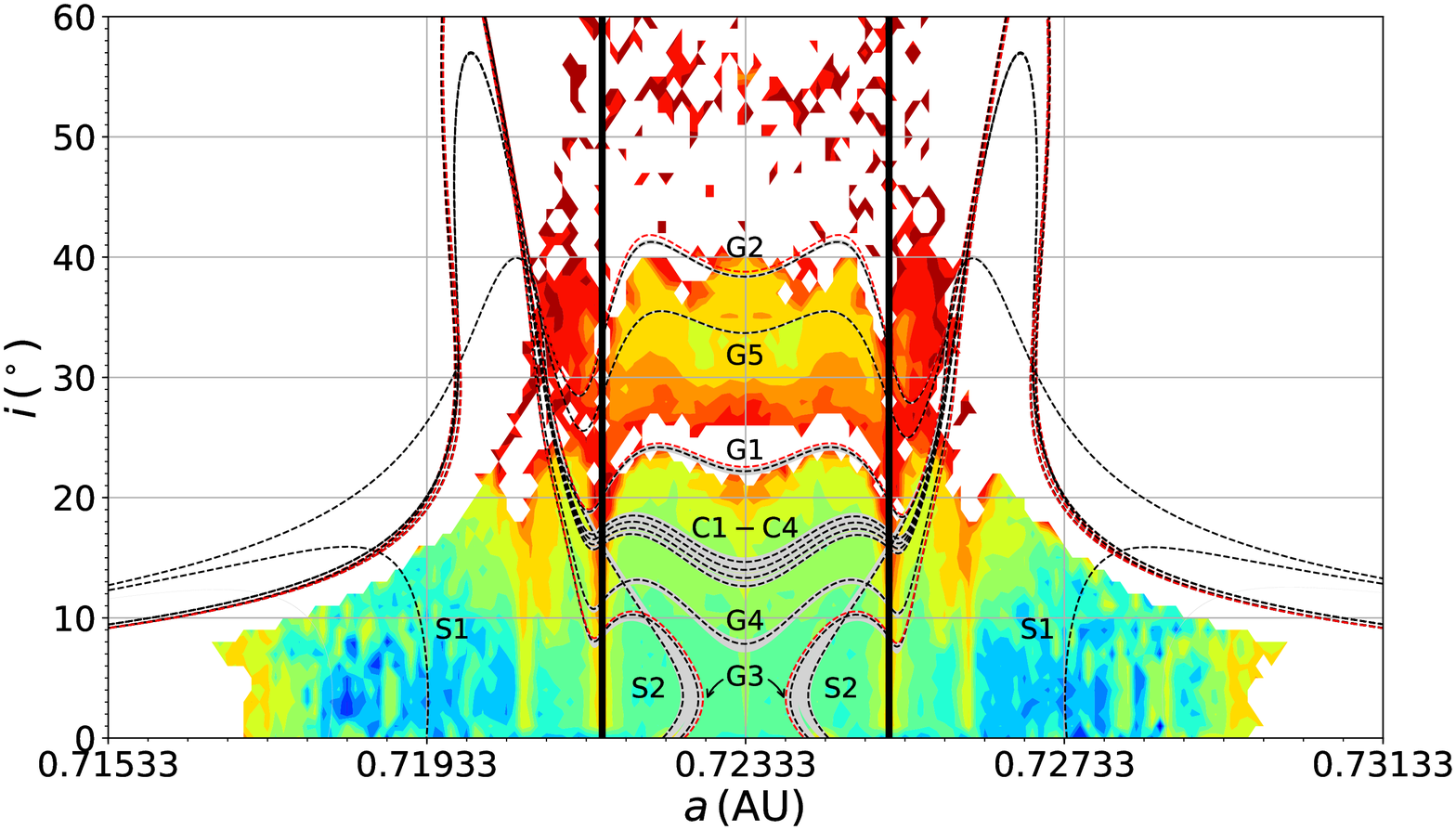}
	\caption{Secular resonances over the short-time ($1.34\times10^6$ yr) dynamical map on the $(a_0,i_0)$ plane (see text for the explanations of the labels along the curves). The \textit{left panel} depicts the linear secular resonances while the \textit{right panel} depicts the fourth-degree secular resonances. The tadpole and horseshoes orbits are separated by the vertical dark lines. Grey shadows delimit the coverage areas of the resonances considering the frequency drift of the inner planets. The distinguishable locations of secular resonances deduced from the model including the general relativity are indicated by the red dashed lines (see text).}
	\label{fig:resweb}
\end{figure*}

The major secular resonances detected are shown in Fig.~\ref{fig:resweb}. As we can see, the boundary between the tadpole and horseshoe orbits separates two regimes differently influenced by the secular resonances. Since the values of proper frequencies of horseshoe orbits are much larger than those of tadpole orbits, Equation~\eqref{eq:res} involving these frequencies are less sensitive to different combinations of proper frequencies of the planets. Therefore, the corresponding secular resonances are much closer to each other in the horseshoe regime, which means the resonance overlapping is easier to take place. It plays an important role in shaping the boundary of the stability region.

Obviously, the apsidal secular resonances with the Earth ($\nu_3$) and Mars ($\nu_4$) clear the moderate-inclined VTs, as we mentioned before. The locations are consistent with the result from \citetads{2005AJ....130.2912S} although they depict them on the plane of proper orbital elements. This kind of apsidal resonances could increase the eccentricity of VTs to make them planet-crossing \citepads[see e.g.][]{1999ssd..book.....M}. And the orbits dominated by the $\nu_3$ and $\nu_4$ range from $17^\circ$ to $35^\circ$ \citepads{2002MNRAS.334..241B}. The $\nu_2$ secular resonance which is located in the high inclination regime could induce chaos, together with the $\nu_{11}$, which has also been determined by \citetads{2002MNRAS.334..241B} and \citetads{2005AJ....130.2912S}. Actually, orbits over $40^\circ$ could also obtain large eccentricity \citepads{1999Icar..137..293N, 2016MNRAS.460..966G} via von Zeipel-Lidov-Kozai mechanism \citepads{1910AN....183..345V, 1962P&SS....9..719L, 1962AJ.....67..591K, 2019MEEP....7....1I}. The apsidal secular resonance with Saturn ($\nu_6$) lies in the low-inclined stability region. It seems to be of weak strength as Saturn is too far away from VTs. Nodal secular resonances such as $\nu_{13}$ and $\nu_{14}$ could control the variation of inclination \citepads[see e.g.][]{1999ssd..book.....M}. These two resonances could cause a variation of inclination up to $\sim20^\circ$ while inclinations of most of other orbits vary within $10^\circ$. 

Besides the linear secular resonances, we also search the fourth-degree secular resonances. The resonances only involving the apsidal ($g$) and nodal precession ($s$) are denoted by ``G'' and ``S'' type respectively. The rest are labeled as ``C'' type in the name of ``combined''. We show the results in the right panel of Fig.~\ref{fig:resweb}, and the labels in the figure are explained as follows:
\begin{equation}
\label{eqn:gsres}
\begin{aligned}
{\rm G1}:  &  g+g_1-g_2-g_4=0,  &
{\rm G2}:  &  g+g_1-2g_2=0, \\
{\rm G3}:  &  g+g_1-2g_4=0, &
{\rm G4}:  &  g+g_2-2g_4=0, \\
{\rm G5}:  &  2g-g_2-g_4=0, \\
{\rm S1}:  &  s+s_2-2s_6=0, &
{\rm S2}:  &  2s-s_2-s_3=0, \\
{\rm C1}:  &  g-s-(g_3-s_4)=0, &
{\rm C2}:  &  g-s-(g_4-s_4)=0, \\
{\rm C3}:  &  g-s-(g_3-s_3)=0, &
{\rm C4}:  &  g-s-(g_4-s_3)=0.
\end{aligned}
\end{equation}	
Actually, there are so many such resonances that we can only show some representatives of them. Many resonances could share the similar locations in the phase space such as $\nu_2$ and G2, $\nu_6$ and G4. These high-order secular resonances contribute to the fine structures of the dynamical maps. We note that C1--C4, four resonances of the same type ``$g-s$'', gather around $i_0=15^\circ$. Obviously, they give rise to the instability strip there in the dynamical map, as we mentioned in Sect.~\ref{subsec:dymap}. Involving both the apsidal and nodal precession, C1--C4 are supposed to have remarkable influence on both the eccentricity and inclination. By inspecting the orbital evolution, we confirm the enhancement of variations of the eccentricity and inclination for orbits around $15^\circ$. We also note that the vertical strips along the inclination in the dynamical map are supposed to be involved with the ``S'' type resonances.

As well known, the proper frequencies of the terrestrial planets could change over time \citepads{1989Natur.338..237L,1990Icar...88..266L}. The direct result of such ``frequency drift'' is that the related resonances could sweep a certain area in the phase space, affecting more orbits and causing more resonance overlap. This is especially true for C1--C4 since they are close to each other and the frequency drift for the Earth and Mars is particularly remarkable (see Table~\ref{tab:fredrift}). So is for G3 and S2. We note that the values of the frequency drift are calculated from our 1~Gyr integration of the Solar system and certainly they could get larger for a longer integration time. 

\begin{table}[htbp]
	\caption{The mean values ($\bar{\nu}$) and amplitudes ($\max-\min$) of the proper frequencies of terrestrial planets. The data are computed from our 1\,Gyr integration of the Solar system.}
	\centering
	\begin{tabular}{crr|crr}
		\hline
		{} &  $\bar{\nu}$~~~ & $\Delta\nu$~~~ & {}  & $\bar{\nu}$~~~ & $\Delta\nu$~~~ \\
		\hline
		$g_1$ & $40.4172$  & $1.1173$ & $s_1$ & $-42.5867$  & $1.1270$ \\
		$g_2$ & $56.6149$  & $0.1210$ & $s_2$ & $-52.8899$  & $1.6284$ \\
		$g_3$ & $133.2022$ & $1.6949$ & $s_3$ & $-146.4151$ & $2.0098$ \\
		$g_4$ & $137.6797$ & $1.5983$ & $s_4$ & $-137.3604$ & $2.7047$ \\
		\hline
	\end{tabular}
	\tablefoot{The frequencies are given in $10^{-7}\,2\pi\,{\rm yr}^{-1}$. The mean frequencies $\bar{\nu}$ are very close to the values given in \citetads{1990Icar...88..266L}, while the variations $\Delta\nu$ are larger in this table because a longer integration time is adopted here (1~Gyr vs. 200~Myr). }
	\label{tab:fredrift}
\end{table}

Besides the frequency drift caused by the nonlinear gravitational perturbation among planets, the general relativity (GR) also brings modifications to the fundamental frequencies of the inner planets. For example, it produces an extra perihelion precession of 38 arcsec per century to Mercury. This change may then influence the effects of the above mentioned resonances. 

Adopting the first-order post-Newtonian expansion \citepads[e.g.][]{ppn-standish}, we include GR effect from the Sun on the planets and VTs in our simulations. New locations of the resonances are obtained after taking into account GR, as shown in Fig.~\ref{fig:resweb}. Since GR has the greatest influence on the apsidal precession of Mercury, the most obvious shift occurs in resonances involving $g_1$, like G1--G3. But still, as we can see from Fig.~\ref{fig:resweb}, GR could only make little difference. In practice, GR is almost ignorable.  

\begin{figure}[htbp]
	\centering
	\resizebox{\hsize}{!}{\includegraphics{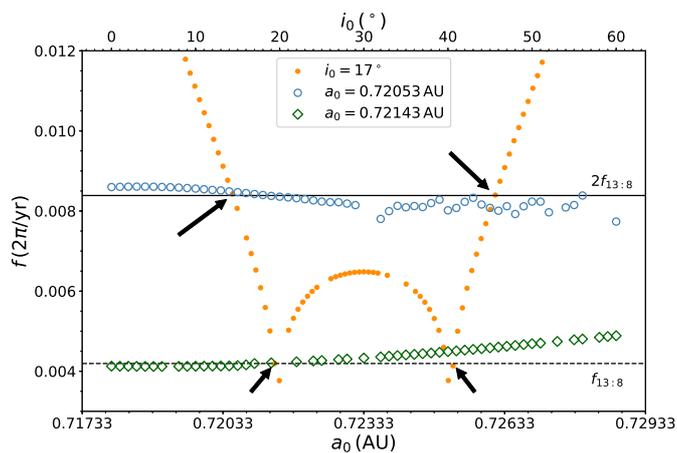}}
	\caption{Libration frequency $f$ of the 1:1 resonant angle for VTs ($\cos\sigma$) versus initial semimajor axis (bottom abscissa) and initial inclination (top abscissa). Solid squares in orange indicate the orbits with varying semi-major axis but fixed initial inclination $i_0=17^\circ$; while open circles in blue and diamonds in green indicate the orbits with varying initial inclination but fixed semi-major axes $a_0=0.72053$ and $0.72143$\,AU, respectively. The frequency of the 13:8 near MMR between the Earth and Venus $f_{13:8}$ and its double value $2f_{13:8}$ are depicted by the dashed and solid lines respectively. The arrows point to the intersections of $f$ with these two lines.}
	\label{fig:secres}
\end{figure}

The secondary resonances could also play a role in Trojan dynamics \citepads[see e.g.][]{2003MNRAS.345.1091M,2006MNRAS.372.1463R,2009MNRAS.398.1217Z,2011MNRAS.410.1849Z,2020A&A...633A.153Z}. The 13:8 near-mean-motion resonance (NMMR) between the Earth and Venus \citepads{2010CeMDA.107...63B} might have a remarkable influence on the orbital behaviours of Venus Trojans. According to our calculation, the angle $13\lambda_3-8\lambda_2$ circulates with a period about 238\,yr, corresponding to a frequency of $f_{13:8}=4.193\times10^{-3}\,2\pi\,{\rm yr}^{-1}$. Secondary resonances take place when the frequency of the 13:8 NMMR ($f_{13:8}$) intersects with the libration frequency of the resonant angle $\cos\sigma$ of VTs ($f$). To determine the locations of secondary resonances on the $(a_0,i_0)$ plane, we may fix one $i_0$ and calculate the frequency $f$ of the resonant angle for varying $a_0$. Along such a straight line of fixed $i_0$ on the $(a_0,i_0)$ plane, the frequency $f$ changes with $a_0$ and then we can check whether $f$ will meet $f_{13:8}$, where a secondary resonance happens. Of course, we can also fix $a_0$ and calculate the variation of $f$ with respect to varying $i_0$. Several examples are shown in Fig.~\ref{fig:secres}.

As we can see from the dynamical spectrum (Fig.~\ref{fig:secres}), for VTs with $i_0=17^\circ$ (arbitrarily chosen as an example), $f=f_{13:8}$ occurs around $a_0=0.72143$ and 0.72523\,AU (indicated by short arrows), which are very close to the separatrix between tadpole and horseshoe orbits (0.72153 and 0.72513\,AU). Therefore, the overlap enhances the instability and causes a small extension of the instability strip away from the centre. The secondary resonance $f=2f_{13:8}$ appears around 0.72053 and 0.72613\,AU (indicated by long arrows in Fig.~\ref{fig:secres}), corresponding to the locations of two vertical strips in the dynamical map, and implying that this secondary resonance might be responsible for such unstable structure. 

Moreover, for orbits with fixed semi-major axes, we find in Fig.~\ref{fig:secres} (open circles and diamonds) that the frequency $f$ of resonant angle changes only little as the inclination increases, implying that the structure caused by the secondary resonances shall appear as vertical strips in the dynamical map on the $(a_0,i_0)$ plane. Such vertical structures can be clearly seen in Fig.~\ref{fig:dynmap}, nearly parallel to the separatrix between tadpole and horseshoe orbits.
And we also note that the overlap of these secondary resonances with $\nu_{13}$ and $\nu_{14}$ secular resonances induces a growing instability along the initial inclination (Fig.~\ref{fig:resweb}). an

Additionally, we inspect the frequency spectra of the eccentricity, and find that a small peak around $f_{13:8}$ exists for VTs located in the aforementioned instability strip while not for VTs out of it, also implying that this instability structure might be related with the 13:8 NMMR. However, considering the fact that the 13:8 resonance is of high order, its effect should be quite small, and we cannot completely exclude the possibility that there are other dynamical mechanisms contributing to the formation of these instability structures in the dynamical map and the equality between frequencies mentioned above is just a coincidence.


\subsection{Long-term stability }\label{subsec:longt}

\begin{figure}[htbp]
	\centering
	\resizebox{\hsize}{!}{\includegraphics{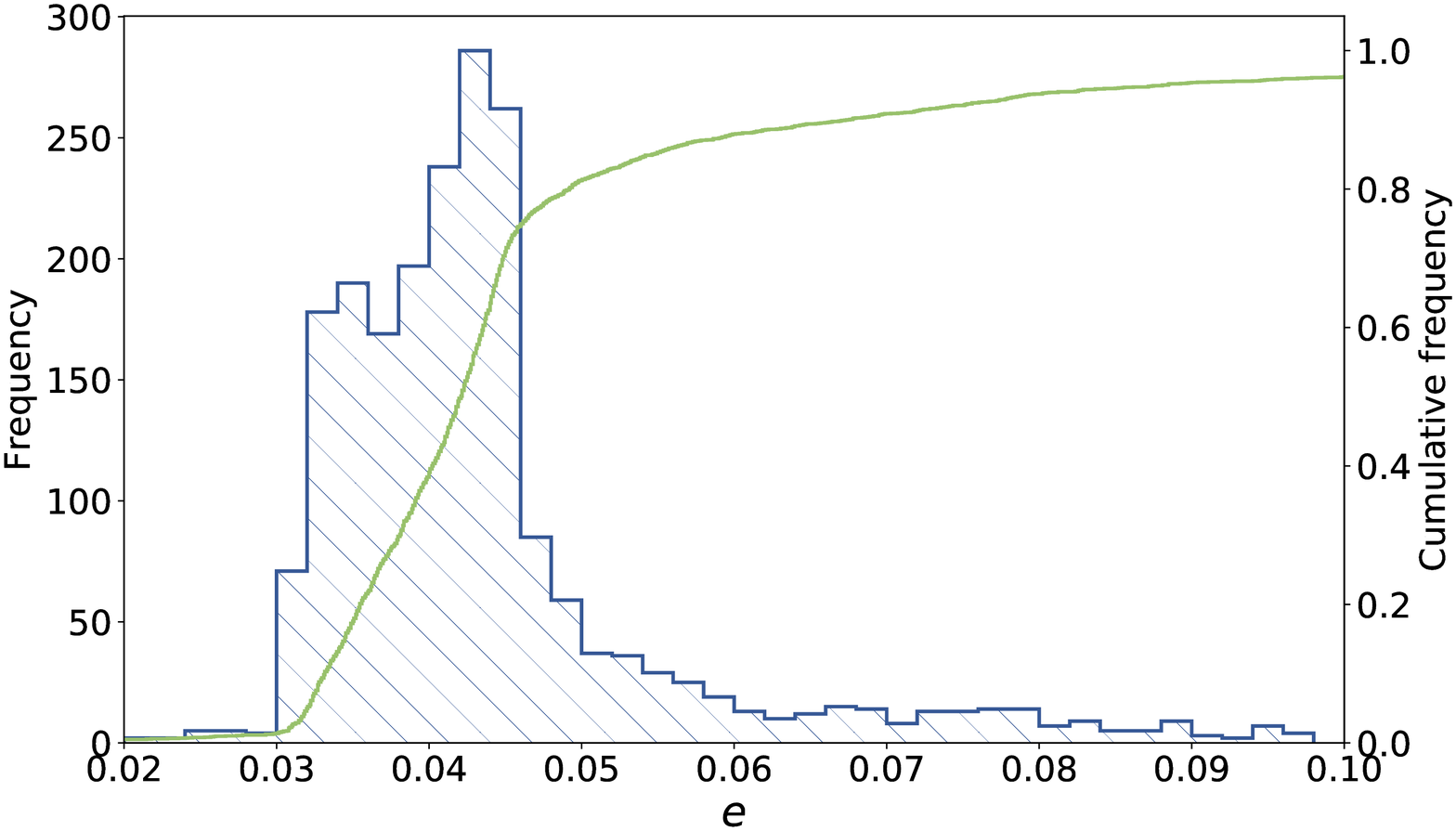}}
	\caption{The distribution of eccentricity of the initial population for the 4.5\,Gyr integration. The blue bins indicate the frequency of $e$ while the green curve is the cumulative frequency. The eccentricity smaller than 0.02 or larger than 0.1 is not shown for a better vision.}
	\label{fig:e0}
\end{figure}

To check the long-term stability of VTs, we set those orbits that survive the $1.07\times10^7$\,yr integration as the initial population (2271 orbits in total) and extend the integration to 4.5\,Gyr using the \emph{Mercury6} \citepads{1999MNRAS.304..793C} integrator package. Since these orbits are the survivors of the $\sim$10\,Myr evolution of the originally near circular orbits, they are reasonably expected to occupy the most stable region in the orbital elements space. Particularly, the instantaneous eccentricities of them may have formed the eccentricity distribution (Fig.~\ref{fig:e0}) of long-term stability. 

As shown in Fig.~\ref{fig:e0}, the instantaneous eccentricities are distributed mainly in between $e=0.03$ and 0.05, with sharp drops on both sides. In fact, this is consistent with the results in  \citetads{2019ApJ...873L..16P} where the most stable orbits surviving the Solar system age have initial eccentricities around $0.04$ (see Fig.~4(a) therein). 

\begin{figure}[htbp]
	\centering
	\resizebox{\hsize}{!}{\includegraphics{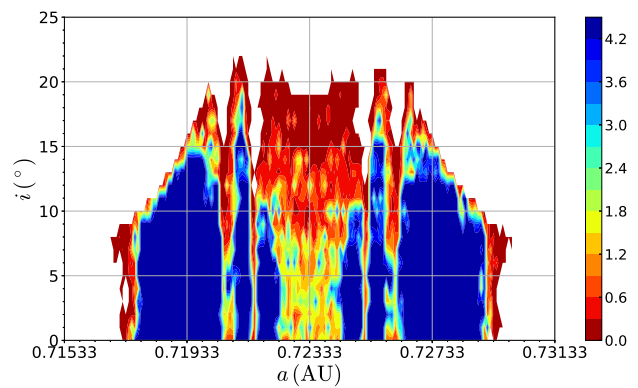}}
	\caption{Lifespans of VTs around the $L_4$ point on the $(a_0,i_0)$ plane. The color indicates the lifespans in gigayear in logarithm. Only the orbits that survive the $1.07\times10^7$\,yr integration are included and orbits over $25^\circ$ are abandoned for a better vision.}
	\label{fig:lf}
\end{figure}

Integrated up to 4.5\,Gyr, the lifespans of these orbits are shown in Fig.~\ref{fig:lf}. About 39.5\% of the orbits could survive the age of the Solar system and most of them are on horseshoe orbits, agreeing very well with the conclusion drawn from the dynamical map (Fig.~\ref{fig:dynmap}).

For tadpole orbits, the long-time integrations confirm the results we obtained via the frequency analysis method (see Sect.~\ref{subsec:fma}). C1--C4 do destabilize the orbits around $15^\circ$, driving them out of the co-orbital region within 1\,Gyr. Resonances including $\nu_6$, S2, G3 and G4 could also contribute to the escape of tadpole orbits, but are less effective than C1--C4. Certainly, VTs on the boundary of the stability region lose their stability in such long integration.

Overall, if we just consider a pure gravitational model of the Solar system, there do exist some primordial Venus companions, especially on horseshoe orbits. \citetads{2005AJ....130.2912S} found no stable orbits that could survive 4.5\,Gyr, probably owing to a small initial population (30). \citetads{2019ApJ...873L..16P} obtained a surviving ratio of only 8.2\% since there are too many high-eccentric orbits in their initial population while the eccentricity of the initial population in our simulation is almost concentrated around $0.04$. 

Since the SN is an indicator of orbital regularity, its value is correlated negatively with the lifespan of an orbit. We check the SN values (calculated from the $1.34\times 10^6$\,yr integration) and lifespans of these orbits and find that 95\% of orbits surviving 4.5\,Gyr have $\log_{10}\text{SN}\le 2.06$. Thus, we regard $\text{SN}=10^{2.06}$ as the critical value for the most stable orbits, and mark the most stable areas in Fig.~\ref{fig:dynmap_e} by the contour curve. Apparently, the stable regions can hardly extend to $e_0=0.1$ and most of them are in the low eccentricity ($e\le 0.05$) region. Therefore, although the instantaneous eccentricities (0.00675 at time zero) of these initial population are very low, their distribution after $\sim$10$^7$\,yr evolution (Fig.~\ref{fig:e0}) agrees basically with both the eccentricities of stable orbits found by \citetads{2019ApJ...873L..16P} and our assumption that long-term stable orbits are mainly low-eccentricity ones. We conclude that the initial population adopted here to examine the long-term stability are representative. Even though some orbits with higher eccentricities (0.05--0.1) could survive 4.5\,Gyr in this pure Newtonian gravitation model, the stability of all these orbits will be destroyed by the Yarkovsky effect finally, as we will show below. 

In addition to the Yarkovsky effect, the general relativity (GR), which may also change the orbital stabilities, is not included in the simulations so far. We will check the influence of GR in greater details in the following section and the combined effects of both GR and Yarkovsky effect will then be investigated subsequently. 

\section{General relativity and Yarkovsky effect}\label{sec:gryark}

\subsection{General relativity}
The effect of GR was hardly ever mentioned in the researches on the dynamical stability of objects in the Solar system, because it was generally thought to be negligible unless in the very close vicinity of the Sun. 
VTs are on the orbit of Venus with a semi-major axis of about 0.7\,AU, so their long-term stability could be affected in a non-negligible level both directly by GR and indirectly by its influences on the motion of planets.

\begin{figure}[htbp]
	\centering
	\resizebox{\hsize}{!}{\includegraphics{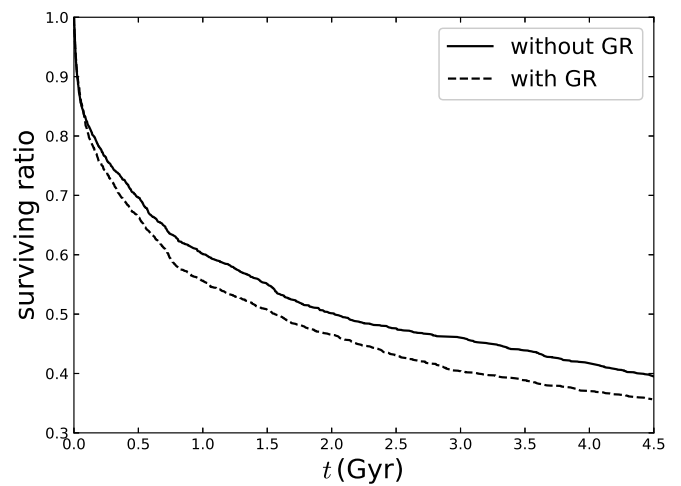}}
	\caption{The surviving ratio of test particles during 4.5\,Gyr.  The dashed line is for the model including GR while the solid line for without GR. 
	}
	\label{fig:sur}
\end{figure}

To investigate how GR affects the long-term stability of VTs, we set the same initial population (2271 test particles surviving $1.07\times10^7$\,yr) and integrate their orbits to 4.5\,Gyr in a model including GR. During the integration, some orbits lose their stability and escape from the Trojan region, and the surviving ratios of test particles during the whole time are illustrated in Fig.~\ref{fig:sur}. For comparison, the surviving ratio in the model without GR is plotted in the same figure. 

Without GR, 1365 test particles survive 1\,Gyr and 898 survive 4.5\,Gyr. Including GR, these two numbers turn to be 1264 and 810, respectively. The relative differences, both at 1\,Gyr and at 4.5\,Gyr, are less than 10\%, implying that GR does affect but not substantially the stability of VTs. 

\begin{figure}[htbp]
	\centering
	\resizebox{\hsize}{!}{\includegraphics{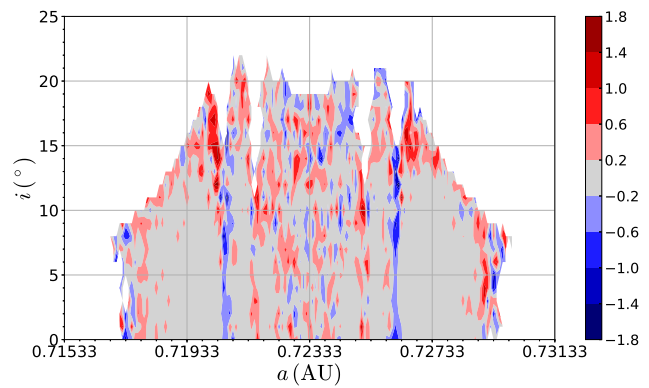}}
	\caption{The effect of GR in VTs' region. The colour bar represents $\log_{10}(T_1/T_2)$ (see text). Thus the grey means that the lifetimes in both model are the same (4.5\,Gyr), while the red indicates that the corresponding orbit is destabilized to some extent by GR ($T_1>T_2$) and the blue is the opposite.
	}
	\label{fig:lt}
\end{figure}

To further show the influence of GR on VTs' orbital stability, we calculate the value of $\log_{10}(T_1/T_2)$ where $T_1$ is the lifetime of a test particle in the model without GR and $T_2$ is that with GR, and plot the result in Fig.~\ref{fig:lt}. As shown in Fig.~\ref{fig:lt}, the red and blue colors, indicating correspondingly the destabilizing and stabilizing effect of GR, mix up in the region. But still, we can find that the red parts (destabilization) mainly locate in the middle of VTs' region, where the resonances involving the apsidal precession of inner planets and VTs gather (see Fig.~\ref{fig:resweb}). The stabilized parts (blue) mainly locate around the secondary resonance $f=2f_{13:8}$ at $a=0.72053$\,AU, $0.72613$\,AU, implying that the secondary resonance is almost destroyed due to GR.

\subsection{Yarkovsky effect}\label{sec:yark}

The Yarkovsky effect produces a recoil force that arises from the absorption and anisotropic re-radiation of thermal energy, resulting in the long-term evolution of the semi-major axis for small objects orbiting the Sun \citepads[for a brief review see e.g.][]{2006AREPS..34..157B}. Due to the rotation and revolution of the object, an asteroid may suffer the diurnal Yarkovsky effect and the seasonal effect. Generally, the diurnal effect is much more remarkable than the seasonal one unless the spin axis is parallel to the orbital plane. Therefore, it is reasonable to consider only the diurnal effect but neglect the seasonal one.

The Yarkovsky effect increases continuously the semi-major axis of an asteroid if it rotates prograde or decreases the semi-major axis if it is a retrograde rotator. But the situation can be complicated when the asteroid is locked in an MMR. For Trojan asteroids, \citetads{2017MNRAS.471..243W} show that the libration amplitude of the semi-major axis undergoes remarkable variation when the Yarkovsky effect works with the 1:1 MMR together. The libration amplitude of semi-major axis decreases with time for prograde rotating objects while it increases for retrograde ones. This phenomenon is observed in the numerical simulations of the Earth Trojans \citepads{2019A&A...622A..97Z} and Jupiter Trojans \citepads{2019A&A...630A.148H}. Due to this mechanism, Trojans are pushed either towards or away from the center of libration, neither of which leads to long-term stable region, because the central part of tadpole orbits is less stable (see Fig.~\ref{fig:dynmap} and Fig.~\ref{fig:lf}), just as in the case of the Earth Trojans \citepads{2019A&A...622A..97Z}.

The Yarkovsky effect on an asteroid is determined in a complicated way by its distance to the Sun, its  thermal parameters, size, bulk density, surface density, spin rate, and the obliquity of its spin axis. Unfortunately, these parameters generally are poorly determined. 
Following \citetads{2005AJ....130.2912S} and \citetads{2013Icar..223..844B}, we may parameterize the ``typical'' VTs with a bulk density of 2.5\,g\,cm$^{-3}$, a surface density of 1.5\,g\,cm$^{-3}$, a surface thermal conductivity of 0.001\,W\,m$^{-1}$\,K$^{-1}$, a specific heat capacity of 680\,J\,kg$^{-1}$\,K$^{-1}$, an emissivity of 0.9, and a Bond albedo of 0.1. For typical rotators in the Solar system, as mentioned in \citetads{2006AREPS..34..157B}, an object with a radius $R=1$\,km has a rotation period of 5000\,s, and the rotation period is proportional to the radius.
Based on these parameters, we simplify the formula of the semi-major axis drift rate for asteroids around Venus orbit due to the diurnal Yarkovsky effect as \citepads{1999A&A...344..362V, 2020MNRAS.493.1447X} 
\begin{equation} \label{eq:yark}
\dot{a}_Y=3274\cdot\frac{\cos\gamma}{R^{3/2}}\,\text{AU/Gyr},
\end{equation}
where $\gamma$ represents the obliquity of the spin axis, and $R$ is the asteroid radius given in meter. 

\subsection{Stability under Yarkovsky effect}
We perform numerical simulations of the motion of VTs in a model including the Yarkovsky effect that drives the semi-major axis of the Trojan to migrate following the rule in Equation~\eqref{eq:yark}. We select from previous simulations those 1365 test particles that survive 1\,Gyr without GR and 1264 test particles surviving 1\,Gyr with GR, and then integrate their orbits to 4.5\,Gyr under the Yarkovsky effects of different magnitudes. 
Several Yarkovsky drift rates $\dot{a}_Y=\pm0.05$, $\pm0.25$, $\pm0.5$ and $\pm$2\,AU/Gyr are adopted. Assuming that the spin axis is perpendicular to the orbital plane, i.e. $\gamma=0^\circ$ or $180^\circ$, the Yarkovsky drift rates chosen above can be translated to asteroid sizes of radii of 1600, 560, 350 and 140\,m, respectively, according to Equation~\eqref{eq:yark}.

\begin{figure}[htbp]
	\centering
	\resizebox{\hsize}{!}{\includegraphics{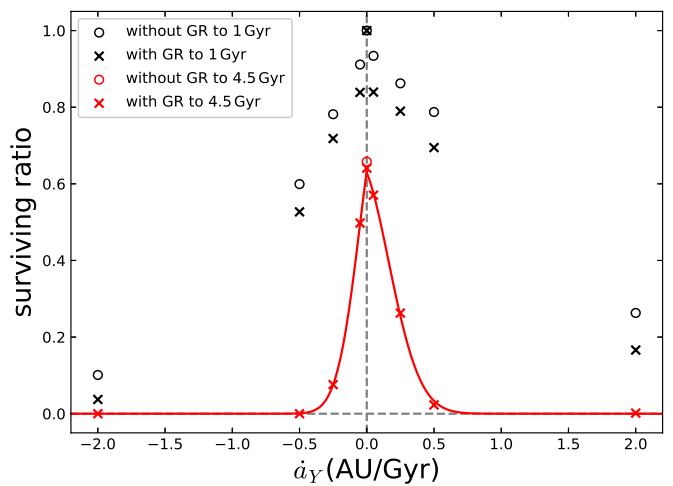}}
	\caption{The surviving ratio of VTs under Yarkovsky effect. The black color refers to the results of integration to 1\,Gyr, while the red for 4.5\,Gyr. The crosses and open circles refer to the results with and without GR, respectively. The surviving ratio of 1.0 right at the center is surely for the case $\dot{a}_Y=0$ (without Yarkovsky effect). Solid lines are fitting curves of the data for 4.5\,Gyr with GR (red crosses, see text).}
	\label{fig:yark}
\end{figure}

The results of our simulations are summarized in Fig.~\ref{fig:yark}, in which the surviving ratios of test Trojans at the end of each simulations are plotted. 

Apparently, the Yarkovsky effect decreases the surviving ratio. As the strength of Yarkovsky effect increases, the surviving ratio drops down sharply.
Similar relationship has been observed for the Earth Trojans influenced by the Yarkovsky effect \citepads{2013CeMDA.117...91M,2019A&A...622A..97Z}.
Meanwhile, just as in the case of Earth Trojans, the asymmetry of stability that the retrograde spinning Trojans ($\dot{a}_Y<0$) are less stable than the prograde spinning ones ($\dot{a}_Y>0$) with the same Yarkovsky effect (i.e. the same $|\dot{a}_Y|$), can be seen here in Fig.~\ref{fig:yark} too. For the same $|\dot{a}_Y|$ value, the surviving ratio of the inward-migrating retrograde rotators ($\gamma=180^\circ$) is lower than that of the outward-migrating prograde rotators ($\gamma =0^\circ$). This happens because of the different responses of the libration amplitude of the Trojans to the inward or outward semi-major axis drift \citepads{2017MNRAS.471..243W,2019A&A...622A..97Z,2019A&A...630A.148H}, as well as because of the different stability structures in the inner and outer regions of the dynamical map (Fig.~\ref{fig:dynmap} and Fig.~\ref{fig:lf}).

Comparing the results taking into account the effect of GR (crosses) and the results without GR (open circles) in Fig.~\ref{fig:yark}, we may find that the influence of GR becomes more prominent than that shown in Fig.~\ref{fig:sur}, implying that the effect of GR on destabilizing Trojans' motion is enhanced after cooperating with the Yarkovsky effect. In fact, the destabilization of GR happens mainly either in the inner central region or around the boundary of the stable island (red parts of Fig.~\ref{fig:lt}), into which Trojans will always be pushed by the Yarkovsky effect in some certain time. As a result, the Yarkovsky effect amplifies the effect of GR in a sense.

According to our experiences obtained in dealing with the Earth Trojans \citepads{2019A&A...622A..97Z}, the relationship between the surviving ratio and the Yarkovsky drifting rate shown in Fig.~\ref{fig:yark} can be numerically fitted by piecewise Gaussian function for the positive and negative $\dot{a}_Y$, separately. The fitting curves for test particles surviving 4.5\,Gyr with GR are plotted in Fig.~\ref{fig:yark} (red solid lines), and they can be used to impose  constraints on the sizes of primordial VTs that probably still exist today.
If a surviving ratio of 1\% is taken as the cut-off probability of existence \citepads[as in][]{2019A&A...622A..97Z}, which corresponds to  $\dot{a}_Y=-0.50$\,AU/Gyr and $+0.77$\,AU/Gyr on the fitting curves for retrograde and prograde spinning asteroids respectively, we may infer from Equation~\eqref{eq:yark} that a primordial VT surviving the Yarkovsky effect for 4.5\,Gyr cannot be smaller than 350\,m in radius if it's a retrograde rotator ($\gamma=180^\circ$) and it should be larger than 260\,m for a prograde rotator ($\gamma=0^\circ$).  

Overall, due to the effect of GR and the Yarkovsky effect, the probability of the existence of primordial VTs that avoid being detected today is extremely low, which implies that the dust ring associated with Venus should be replenished by some other sources.

\section{Conclusion and Discussion}\label{sec:disc}

The orbital stability of Venus Trojans (VTs) is thoroughly investigated via numerical simulations in this paper. 

In a full gravitational model consisting of the Sun and eight planets, the motions of thousands of test particles in the region of 1:1 MMR with Venus are integrated to $\sim$10\,Myr. Based on these simulated orbits and using the spectral number as the indicator of orbital regularity, we portray detailed dynamical maps on the $(a_0,i_0)$ plane. Stable regions of low-inclination $(\lesssim 15^\circ)$ for VTs mainly on horseshoe orbits are found (Fig.~\ref{fig:dynmap}). Most of them are located in the region of low-eccentricity $(\lesssim 0.05)$ (Fig.~\ref{fig:dynmap_e}).

Using the frequency analysis method, we determine the proper frequencies of VTs and then locate the resonances that contribute to shaping the structures of the dynamical maps. We find that the instability of motion is mainly caused by the linear apsidal secular resonances $\nu_2, \nu_3, \nu_4$ and $\nu_6$, while the nodal secular resonances $\nu_{13}$ and $\nu_{14}$ may excite the inclination variation up to $\sim$20$^\circ$ (Fig.~\ref{fig:resweb}). The secondary resonances related to the frequency of the quasi 13:8 MMR between the Earth and Venus are likely to be responsible for the narrow vertical strips of instability on the $(a_0,i_0)$ plane (Fig.~\ref{fig:secres}).
Some higher-order secular resonances involving the apsidal and nodal precession of Trojans, have been located as well, and they produce the fine structures in the dynamical maps. Their effects are further enhanced by the frequency drift (of planets' precession) in the inner Solar system, and they are responsible partly for the depletion of Trojans on the tadpole orbits in long term (Fig.~\ref{fig:resweb}). Compared to the effect of frequency drift, the general relativity introduces only tiny changes to the stability (Fig.~\ref{fig:sur}). In summary, in this ``pure gravitational'' model, either with or without the general relativity, VTs of low inclinations $(\lesssim 15^\circ)$ and small eccentricities $(\lesssim 0.05)$ (Fig.~\ref{fig:e0}), particularly those on horseshoe orbits, may survive the Solar system age (Fig.~\ref{fig:lt}).  

However, the stability of asteroids in the Trojan region will be destroyed by the Yarkovsky effect (Fig.~\ref{fig:yark}). As shown in our previous work \citepads{2019A&A...622A..97Z}, the Yarkovsky effect has depleted practically all primordial Earth Trojans (ETs, if there were) that are so small that they could avoid being detected by the surveys like OSIRIS-REx. Assuming that VTs have the same (typical) physical and thermal parameters as ETs, the Yarkovsky effect on VTs must be stronger than that on ETs because they are closer to the Sun. Taking a surviving ratio of 1\% as the critical probability, our calculations show that a retrograde spinning VT surviving the Yarkovksy effect for 4.5\,Gyr cannot be smaller than 350\,m (or 260\,m if it's spinning prograde). In other words, those asteroids smaller than 350\,m (or 260\,m) in radius have a probability of $<1\%$ to survive the Solar system age. 

Meanwhile, recent surveys dedicated to searching for VTs put upper limit on the sizes of these hypothetical objects. The null detection of VTs by the Zwicky Transient Facility (ZTF) twilight survey \citepads{2020AJ....159...70Y} with limiting $r$-band magnitude 19.5 suggests that there is no VT larger than 250\,m-750\,m (assuming albedos between 0.45 and 0.04). And the Cerro-Tololo Inter-American Observatory (CTIO) twilight survey \citepads{2020PSJ.....1...47P} with limiting magnitude 21 (corresponding to radius $\sim$200\,m for S-type and $\sim$500\,m for P-type asteroid) reports no detection of VTs either. 

In one word, the Yarkovsky effect expels small VTs while the observational surveys deny the existence of large ones. Hence, we may conclude that the undetected primordial VTs are very unlikely to exist today. 

Our calculations suggest that the dust ring around the orbit of Venus cannot be the by-product of erosion of primordial VTs. Still, temporary capture of asteroids by the resonance may be possible and could be the source for co-orbital dust on smaller time scales. There are currently five known co-orbital objects of Venus that are small asteroids on chaotic orbits. Their estimated lifetimes in the 1:1 MMR with Venus are rather short: object 2001 CK32 is currently in transition between horseshoe and quasi-satellite orbit with $e\simeq0.38$, and numerical studies suggest the resonant capture last for less than 10\,000 years \citepads{2004Icar..171..102B}. Asteroid 2002 VE68 currently stays on a fairly chaotic quasi-satellite orbit with $e\simeq0.41$ and the temporary capture time is of thousands of years \citepads{2004MNRAS.351L..63M}.  The three other co-orbital asteroids 2012 XE133, 2013 ND15 and 2015 WZ12 \citepads{2013MNRAS.432..886D, 2014MNRAS.439.2970D, 2017RNAAS...1....3D} are much fainter, and on eccentric thus less stable orbits. Because of their short lifetimes and short trajectory arcs in the vicinity of Venus orbit, the contribution from these temporary co-orbital asteroids to the dust ring - if at all - can only be seen as a temporary phenomenon.

An alternative mechanism to maintain the dust ring is a continuous concentration and dispersion of dust particles in the 1:1 orbital resonance with Venus. Due to their small size, dust particles receive significant nongravitational perturbations, such as the radiation pressure, Poynting-Robertson drag, and solar wind drag \citepads{1979Icar...40....1B}. Driven by these perturbations, dust particles migrate toward the Sun from their birthplaces (main belt asteroids or Jupiter family comets), and the orbital resonances with planets may trap or delay the traveling particles so that dust rings can form.
 
It is known that the capture of small objects into external resonances takes place, but it does not happen for internal resonances \citepads{1993CeMDA..57..373S, 1994Icar..110..239B,1995Icar..113..403L}. The co-orbital resonance, being neither an external nor an internal resonance, is special. Numerical studies already indicate the low capture probability of dust in the external and 1:1 MMRs with Venus \citepads{1989Natur.337..629J, 1994Natur.369..719D, 2012LPICo1667.6201J}. Recently, \citetads{2020A&A...635A..10S} study in great detail the formation of resonant dust rings in the inner Solar system,  and they find that the neighboring planet (Earth here) severely impairs the ability of the external resonances of Venus to trap dust particles. Consequently, circumsolar rings at the external and the co-orbital resonances with Venus cannot be formed in this way. 

Even the dust particles have been captured, they may escape from the resonances with the aid of the nongravitational perturbations. \citetads{2021A&A...645A..63Z} find that the nongravitational perturbations shift significantly the location and size of the stability islands around the Lagrange points of Venus, resulting in short lifespans of particles in the co-orbital region. After considering the interaction of charged dust grains with the interplanetary magnetic field, they find that the Lorentz force further destabilizes the orbits.   

Therefore, the primordial VTs, the temporary co-orbital asteroids and the migrating dust cloud, none of them is adequate to explain a long-standing dust ring around the orbit of Venus. Either there are some other sources that we do not know yet, or the dust ring itself is only a temporary phenomenon. The origin of the dust ring associated with Venus is still an open question. 


\begin{acknowledgements}
We thank the anonymous referee for the insightful comments that help us to improve the manuscript. This work has been supported by the National Key R\&D Program of China (2019YFA0706601) and National Natural Science Foundation of China (NSFC, Grants No.11933001 \& No.12150009). We also acknowledge the science research grants from the China Manned Space Project with NO.CMS-CSST-2021-B08. CL was supported by the FWF national science fund with project number P-30542.
\end{acknowledgements}

%
%


   

\end{document}